\documentclass{PoS}
\usepackage{graphicx}
\usepackage{amsmath}
\usepackage{bm}
\usepackage[numbers,sort&compress]{natbib}
\newcommand{\be}{\begin{equation}}
\newcommand{\en}{\end{equation}}
\newcommand{\bea}{\begin{eqnarray}}
\newcommand{\ena}{\end{eqnarray}}
\newcommand{\lbl}[1]{\label{eq:#1}}
\newcommand{\lbltab}[1]{\label{tab:#1}}

\newcommand{\rf}[1]{(\ref{eq:#1})}
\newcommand{\Table}[1]{\ref{tab:#1}}
\newcommand{\fig}[1]{\ref{fig:#1}}

\newcommand{\outleft}{%
\mathrel{\setbox1=\hbox{$\scriptstyle{out}$}\setbox0=\hbox{$\langle$}\copy0\kern-0.3\wd0\lower1.1\ht0\copy1\kern-0.4\wd1}}

\newcommand{\bc}{\begin{center}}
\newcommand{\ec}{\end{center}}
\newcommand{\bt}{\begin{tabular}}
\newcommand{\et}{\end{tabular}}
\newcommand{\ba}{\begin{array}}
\newcommand{\ea}{\end{array}}
\newcommand{\gapprox}{%
\mathrel{%
\setbox0=\hbox{$>$}\raise0.6ex\copy0\kern-\wd0\lower0.65ex\hbox{$\sim$}}}
\newcommand{\lapprox}{%
\mathrel{%
\setbox0=\hbox{$<$}\raise0.6ex\copy0\kern-\wd0\lower0.65ex\hbox{$\sim$}}}
\renewcommand{\slash}[1]{%
\mathrel{\setbox0=\hbox{$/$}\copy0\kern-\wd0\hbox{$#1$}}}
\newcommand{\im}{{\rm Im\,}}

\newcommand{\meta}{m_\eta}

\newcommand{\mpi}{m_\pi}
\newcommand{\mpid}{m_\pi^2}
\newcommand{\fpid}{F_\pi^2}
\newcommand{\mkd}{m_K^2}

\newcommand{\disc}{\hbox{disc}}

\newcommand{\Kbar}{\bar{K}}

\makeatletter
\newcommand*\if@single[3]{%
  \setbox0\hbox{${\mathaccent"0362{#1}}^H$}%
  \setbox2\hbox{${\mathaccent"0362{\kern0pt#1}}^H$}%
  \ifdim\ht0=\ht2 #3\else #2\fi
  }
\newcommand*\rel@kern[1]{\kern#1\dimexpr\macc@kerna}
\newcommand*\widebar[1]{\@ifnextchar^{{\wide@bar{#1}{0}}}{\wide@bar{#1}{1}}}
\newcommand*\wide@bar[2]{\if@single{#1}{\wide@bar@{#1}{#2}{1}}{\wide@bar@{#1}{#2}{2}}}
\newcommand*\wide@bar@[3]{%
  \begingroup
  \def\mathaccent##1##2{%
    \if#32 \let\macc@nucleus\first@char \fi
    \setbox\z@\hbox{$\macc@style{\macc@nucleus}_{}$}%
    \setbox\tw@\hbox{$\macc@style{\macc@nucleus}{}_{}$}%
    \dimen@\wd\tw@
    \advance\dimen@-\wd\z@
    \divide\dimen@ 3
    \@tempdima\wd\tw@
    \advance\@tempdima-\scriptspace
    \divide\@tempdima 10
    \advance\dimen@-\@tempdima
    \ifdim\dimen@>\z@ \dimen@0pt\fi
    \rel@kern{0.6}\kern-\dimen@
    \if#31
      \overline{\rel@kern{-0.6}\kern\dimen@\macc@nucleus\rel@kern{0.4}\kern\dimen@}%
      \advance\dimen@0.4\dimexpr\macc@kerna
      \let\final@kern#2%
      \ifdim\dimen@<\z@ \let\final@kern1\fi
      \if\final@kern1 \kern-\dimen@\fi
    \else
      \overline{\rel@kern{-0.6}\kern\dimen@#1}%
    \fi
  }%
  \macc@depth\@ne
  \let\math@bgroup\@empty \let\math@egroup\macc@set@skewchar
  \mathsurround\z@ \frozen@everymath{\mathgroup\macc@group\relax}%
  \macc@set@skewchar\relax
  \let\mathaccentV\macc@nested@a
  \if#31
    \macc@nested@a\relax111{#1}%
  \else
    \def\gobble@till@marker##1\endmarker{}%
    \futurelet\first@char\gobble@till@marker#1\endmarker
    \ifcat\noexpand\first@char A\else
      \def\first@char{}%
    \fi
    \macc@nested@a\relax111{\first@char}%
  \fi
  \endgroup
}
\makeatother

\newcommand{\Kp}{{K^+}}

\newcommand{\pip}{{\pi^+}}
\newcommand{\pim}{{\pi^-}}
\newcommand{\piz}{{\pi^0}}
\newcommand{\alphaz}{{\alpha_0}}
\newcommand{\betaz}{{\beta_0}}
\newcommand{\gammaz}{{\gamma_0}}
\newcommand{\betaone}{{\beta_1}}

\title{$a_0-f_0$ mixing in the Khuri-Treiman equations for $\eta\to 3\pi$}

\ShortTitle{$a_0-f_0$ mixing in the Khuri-Treiman}

\author{M. Albaladejo\\
        Instituto de F\'{\i}sica Corpuscular (IFIC), Centro Mixto
  CSIC-Universidad de Valencia, Spain\\
        E-mail: \email{Miguel.Albaladejo@ific.uv.es}
}

\author{\speaker{B. Moussallam}\\
        Groupe de Physique Th\'eorique, IPN (UMR8608), Universit\'e
  Paris-Sud 11, Orsay, France\\
        E-mail: \email{moussall@ipno.in2p3.fr}
}

\abstract{
A reliable determination of the isospin breaking double quark mass ratio
from precise experimental data on $\eta\to 3\pi$ decays should be based on the
chiral expansion of the amplitude supplemented with a Khuri-Treiman type
dispersive treatment of the final-state interactions. We discuss an extension
of this formalism which allows to estimate the effects of the $a_0(980)$ and
$f_0(980)$ resonances and their mixing on the $\eta\to 3\pi$ amplitudes. 
Matrix generalisations of the equations describing elastic $\pi\pi$
rescattering with $I=0,\,2$ are introduced which accomodate both
$\pi\pi/K\Kbar$ and $\eta\pi/K\Kbar$ coupled-channel rescattering.  Isospin
violation induced  by the physical $K^+-K^0$ mass difference and by direct
$u-d$ mass difference effects are both accounted for in the dispersive
integrals. 
Numerical solutions are constructed which illustrate how the large resonance
effects at 1 GeV propagate down to low energies. They remain small in the
physical region of the decay, due to the matching constraints with the NLO
chiral amplitude, but they are not negligible and go in the sense of further
improving the agreement with experiment for the Dalitz plot parameters.     
}

\FullConference{The 8th International Workshop on Chiral Dynamics, CD2015 ***\\
		29 June 2015 - 03 July 2015\\
		Pisa,Italy}

\begin{document}

\section{Introduction}
Precise measurements of Dalitz plot distributions of the $\eta\to
\pi^+\pi^-\pi^0$ (see~\cite{Ambrosino:2008ht}) and $\eta\to
\pi^0\pi^0\pi^0$ decays
(see~\cite{Adolph:2008vn})
have been performed recently (see also the talks by S. Giovannella and S. Fang
at this conference). These measurements provide precious insights into the
workings of the chiral expansion beyond the next-to-leading order (NLO) which
is necessary in order to arrive at a reliable and precise determination of the
isospin breaking double quark mass ratio 
\be
Q^2=\frac{m_s^2-m_{ud}^2}{m_d^2-m_u^2},\quad  m_{ud}=(m_u+m_d)/2\ .
\en 
As an illustration of the importance of NNLO effects, consider the
slope parameter $\alpha$ in the $\eta\to 3\pi^0$ Dalitz plot. The NLO
prediction~\cite{Gasser:1984pr} depends on no coupling constant
(except $F_\pi$)  but fails to agree with experiment, 
\be\ba{l}
\alpha^{NLO}=+1.41\cdot10^{-2}\\ 
\alpha^{exp}=-(3.15\pm0.15)\cdot10^{-2}\\ 
\ea\en
The calculation of the $\eta\to 3\pi$ amplitudes at NNLO has been
performed~\cite{Bijnens:2007pr} but, in this case, predictivity is
limited by the absence of model independent determination of the
relevant coupling constants $C_i^r$.

Part of the $O(p^6)$ (and higher order) chiral effects can be
attributed to final-state interactions (FSI). Indeed, the chiral NLO
amplitude accounts for the FSI at $O(p^2)$ only. The attractive idea
was proposed long ago to treat the FSI part exactly through dispersive
methods while using the chiral expansion in unphysical sub-threshold
regions~\cite{Neveu:1970tn,Roiesnel:1980gd,Kambor:1995yc,Anisovich:1996tx}
and then match the two representations. The formalism proposed by
Khuri and Treiman~\cite{Khuri:1960zz} for treating the $S$-wave FSI in
three-body decays was extended to $P$-wave rescattering and applied to
the $\eta\to 3\pi$ amplitude in
refs.~\cite{Kambor:1995yc,Anisovich:1996tx}. Here, we investigate a
further extension of the KT equations, which goes beyond the elastic
approximation, and thereby includes the influence of the $a_0-f_0$
resonant mixing effects. In this regard, the formalism accounts for
isospin breaking induced by the $K^+-K^0$ mass difference via
unitarity, first considered in ref.~\cite{Achasov:1979xc}, as well as
the direct quark mass matrix effects.
\section{Khuri-Treiman formalism with chiral NLO matching: elastic case}
Khuri and Treiman~\cite{Khuri:1960zz} have shown that the final-state
interaction problem, for three-body decays, can be recast as a problem
of solving sets of integral equations involving functions of one
variable. Its application to the $\eta\to 3\pi$
amplitudes~\cite{Kambor:1995yc,Anisovich:1996tx} involves three
functions, $M_I(w)$, $I=1,2,3$ such that the charged decay amplitude
can be expressed as follows,
\be\lbl{ALdecomp}
{\cal T}^{\eta\to \pi^+\pi^-\pi^0}(s,t,u)= -\epsilon_L\,M(s,t,u)\ ,
\quad \epsilon_L= 
\frac{1}{Q^2}\, \frac{\mkd(\mkd-\mpid)}{3\sqrt3 \fpid\, \mpid}\ 
\en
and
\be
M(s,t,u)=M_0(s)-{2\over3}M_2(s)+(s-u)M_1(t)+M_2(t)+
(t\leftrightarrow u)\  .
\en 
The Mandelstam variables are given by $s=(p_\pip+p_\pim)^2$,
$t=(p_\pim+p_\piz)^2$, $u=(p_\pip+p_\piz)^2$.  Under the assumption of
elastic unitarity, the equations can be expressed in terms of the
Omn\`es functions,
\be
\Omega_I(w)=\exp\Big[\frac{s}{\pi}\int_{4\mpid}^\infty \frac{ds'}{s'(s'-w)}\,
\delta_I(s')
\Big]
\en
where $\delta_I$ is the $\pi\pi$ elastic phase shift with isospin $I$ and
$J=0$ or $1$. The  KT equations, in the four-parameter version proposed in
ref.~\cite{Anisovich:1996tx}, have the following form 
\be\lbl{KTeq}
\ba{l}
M_0(w)= \Omega_0(w)\big[\alpha_0+w\beta_0
+w^2\left(\gamma_0 +\hat{I}_0(w)\right)\big]\\[3pt]
M_1(w)= \Omega_1(w)\,w \big[\beta_1 + \hat{I}_1(w)\big]\\[3pt]
M_2(w)= \Omega_2(w)\,w^2\big[ \hat{I}_2(w)\big]\\
\ea\en
where
\be\lbl{KTint}
\hat{I}_I(w)=-\frac{1}{\pi}\int_{4\mpid}^\infty ds'\,
\frac{\im[1/\Omega_I(s')]}{(s')^n(s'-w)}\,\hat{M}_I(s')\ 
\en
(with $n=2$ when $I=0,2$ and $n=1$ when $I=1$). The functions $\hat{M}_I$,
finally, are the left-cut parts of the $\eta\pi\to (\pi\pi)_I$ partial-wave
amplitudes ${\cal T}_J^I$ with $J=0,1$,
\be\ba{l}
{\cal
  T}_0^0(s)=\frac{\sqrt6\epsilon_L}{32\pi}
\left(M_0(s)+\hat{M}_0(s)\right)\\[3pt] 
{\cal T}_1^1(s)=\frac{\epsilon_L}{48\pi}\kappa(s)
\left(M_1(s)+\hat{M}_1(s)\right)\\[3pt] 
{\cal T}_0^2(s)=-\frac{\epsilon_L}{16\pi}\left(M_2(s)+\hat{M}_2(s)\right)\\
\ea\en
with $\kappa^2(s)=(1-4\mpid/s)(s-(\meta+\mpi)^2)(s-(\meta-\mpi)^2)$.
The functions $\hat{M}_I$ can be expressed in terms of the $M_I$
functions via the representation~\rf{ALdecomp} and performing $J=0,1$
partial-wave projections, such that the equations~\rf{KTeq} form a
linear, self-consistent system. It is easy to verify that these
equations ensure that the partial-wave amplitudes satisfy the correct
elastic unitarity equations,
\be\lbl{Uelastic}
\im[ {\cal T}^I_{J}(s)] = \exp(-i\delta_I(s))\,\sin(\delta_I(s))
{\cal T}^I_{J}(s)\ ,\ J=0,1
\en
Strictly speaking, this form is valid in the unphysical situation where
$m_\eta < 3m_\pi$. The analytical continuation of eq.~\rf{Uelastic} to the
physical case is performed by replacing the imaginary part by the
discontinuity across the unitarity cut (divided by $2i$). Furthermore, in that
case, the complex cut of the functions $\hat{M}_I$ overlaps with the unitarity
cut and these functions diverge at the pseudo-threshold
$s=(m_\eta-m_\pi)^2$. The integrals $\hat{I}_I$ can also be defined (and are
finite) by analytic continuation. All these subtle points are explained in
detail in ref.~\cite{Kambor:1995yc}.

Clearly, the representation~\rf{ALdecomp} is not the most general one for an
amplitude which depends on two independent variables and must therefore hold
only in a restricted region of the Mandelstam variables. It is also easy to
check that this representation implies
\be
\im[ {\cal T}^I_{J\ge2}(s)] = 0
\en
which cannot be exactly correct but represents an acceptable
approximation when  the corresponding $J\ge2$ $\pi\pi$ phase-shifts are
small, that is, in the region $s\lapprox 1$ $\hbox{GeV}^2$.

\subsection{Matching conditions}
The parameters $\alpha_0$, $\beta_0$, $\gamma_0$, $\beta_1$ originate
from the presence of subtractions in the dispersive representations of
the functions $M_I$. These are necessary in order to reduce the
dependence on the integration region $s' > 1$ $\hbox{GeV}^2$. We
consider here a version which leads to four polynomial parameters. It
is particularly convenient, since, in this case, all the parameters
can be fixed from matching with the NLO chiral amplitude. The matching
conditions are obtained from the simple requirement that the
difference between the dispersive and the chiral amplitude of order
$p^n$ should be of chiral order $p^{n+1}$, i.e., in our case,
\be\lbl{Diff}
M^{KT}(s,t,u)-M^{ChPT}(s,t,u)=O(p^6)\ .
\en
This relation is satisfied automatically for the imaginary part of the
difference, which implies that the real part can be expanded as a polynomial
as a function of the variables $s$, $t$, $u$. Equating this polynomial to zero
gives four independent equations. Expressing the chiral amplitude
$M^{ChPT}(s,t,u)$  in the same form as eq.~\rf{ALdecomp} in terms
of three functions $\bar{M}_I$ (see~\cite{Anisovich:1996tx}) these four
matching equations can be written as follows,
\be\lbl{Matcheqs}
\ba{ll}
\alphaz=& 9\,\big(\dfrac{1}{2}{\bar{M}''_2}
        -\hat{I}_2\big)\,s_0^2 
        +3({\bar{M}'_2}
        -  {\bar{M}_1}  )\,s_0
        +  {\bar{M}_0}
        +\dfrac{4}{3}{\bar{M}_2}\\[3pt]
\betaz=&\!\!\!-9\,\big(
         \dfrac{1}{2}{\bar{M}''_2}-\hat{I}_2\big)\,s_0 
                    +{\bar{M}'_0}
                   +3{\bar{M}_1}
        -\dfrac{5}{3}{\bar{M}'_2}
        -\Omega'_0\alphaz\\[0.25cm]
\betaone=&
{\bar{M}'_1}+\dfrac{1}{2}{\bar{M}''_2}-\hat{I}_1-\hat{I}_2\\[3pt] 
\gammaz=& \dfrac{1}{2}{\bar{M}''_0}
         +\dfrac{2}{3}{\bar{M}''_2}
        -\hat{I}_0-\dfrac{4}{3}\hat{I}_2
        -\dfrac{1}{2}\Omega''_0\alphaz-\Omega'_0\betaz
\ea\en
where all functions and their derivatives are to be taken at $w=0$. Note that
the integrals $\hat{I}_I$ carry a linear dependence on the four polynomial
parameters.  

\section{KT equations for $\eta\to 3\pi$ with $K\Kbar$ inelastic channels}
\subsection{Difficulties of the general extension} 
The assumption of elastic unitarity seems well justified for
$\eta\to3\pi$ since the $\pi\pi$ energy $s \le (\meta-\mpi)^2$ and
elastic unitarity is known to hold to a good approximation in the
region $\sqrt{s} < 1$ GeV. However, the KT equations involve integrals
over an infinite energy range. In the Omn\`es integrals, for instance,
there is an arbitrariness as to the choice of the phase to be used
above the $K\Kbar$ threshold.  The subtractions and the matching
equations ensure that the dispersive amplitude in the low energy
region should not depend too much on the choice of phase above 1
GeV. Nevertheless, having a small number of subtractions, one expects that an
improved treatment of the 1 GeV region, should result in better
precision at low energy.  At 1 GeV, the two resonances $a_0(980)$ and
$f_0(980)$ are present and a sharp onset of inelastic $\pi\pi \to
K\Kbar$ scattering in the $S$-wave is observed.

Extending the unitarity relations to include the three
$K\Kbar$ channels leads one to consider the transition amplitudes from
$K^+K^-$, $K^0\bar{K}^0$, $K^+\bar{K}^0$ to $\pi^+\pi^-$, $\pi^0\pi^0$, 
$\pi^+\pi^0$ and to $\eta\pi^0$, $\eta\pi^+$. Each of these physical
amplitudes can be expressed in terms of functions of one variable, analogous
to eq.~\rf{ALdecomp}. Unfortunately, a separation of each amplitude into an
isospin violating and an isospin conserving component is not possible at this
level because the kinematical constraint on $s+t+u$ is different for $K^+K^-$
and $K^0\bar{K}^0$ amplitudes. This separation can be made only after
performing the partial wave projections. This feature complicates the
derivation of a general self-consistent set of KT equations including the
$K\Kbar$ channels. In the following, we will show that a simple
approximation can be made which should provide a sensible order of
magnitude for the influence of these channels on the $\eta\to 3\pi$ amplitude.

\subsection{Coupled-channel unitarity relations}
Let us now focus on the $J=0$
partial waves. We will have to consider the following isospin conserving
amplitudes with $I=0,\,1$
\be
\bm{T}^{(0)}=\begin{pmatrix} \pi\pi\to \pi\pi & \pi\pi\to K\Kbar\\
\pi\pi\to K\Kbar &  K\Kbar\to K\Kbar\\
\end{pmatrix}_{I=0},\quad
\bm{T}^{(1)}=\begin{pmatrix} \eta\pi\to \eta\pi & \eta\pi\to K\Kbar\\
\eta\pi\to K\Kbar &  K\Kbar\to K\Kbar\\
\end{pmatrix}_{I=1}
\en
while for $I=2$, coupling to $K\Kbar$ cannot occur and we will continue to use
the elastic approximation in that case\footnote{Inelasticity will also be
  ignored for $J=1$ scattering. Recall that we are mainly interested here in
  accounting for the effects of the $I=0,1$ scalar resonances.}. 
We can classify the isospin violating
amplitudes into two classes: a) $I=0\to I=1$ transitions and b) $I=1\to I=2$
transitions, \be \bm{T}^{(01)}=\begin{pmatrix} (\pi\pi)_0\to\eta\pi &
(K\Kbar)_0\to \eta\pi\\ (\pi\pi)_0\to (K\Kbar)_1 & (K\Kbar)_0\to (K\Kbar)_1\\
\end{pmatrix},\quad
\bm{T}^{(12)}=\begin{pmatrix} \eta\pi^+\to \pi^+\pi^0\\ K^+\bar{K}^0\to
\pi^+\pi^0\\
\end{pmatrix}
\en 
We can then write the unitarity relations for these two sets of isospin
violating amplitudes which now include $K\Kbar$ inelasticity. For
$\bm{T}^{(01)}$, and to first order in isospin breaking, they read 
\be\lbl{UnitT01}
\im[ {\bm{T}^{(01)}} ]= 
\bm{T}^{(0)*}\,\Sigma^0\, {\bm{T}^{(01)}}
+ {\bm{T}^{(01)*}}\,\Sigma^1\, \bm{T}^{(1)} 
+    \bm{T}^{(0)*}\begin{pmatrix}
0 & 0 \\
0 & {\Delta\sigma_K}\\
\end{pmatrix}\bm{T}^{(1)}\\
\en
with
\be
\Sigma^0(s)=\begin{pmatrix}
\sigma_{\pi\pi}(s) & 0 \\
0 & \sigma_{KK}(s)  \\
\end{pmatrix},\quad
\Sigma^1(s)=\begin{pmatrix}
\sigma_{\eta\pi}(s) & 0 \\
0 & \sigma_{KK}(s)
\end{pmatrix}
\en
and
\be
\sigma_{PQ}(s)=\sqrt{\big(1-{(m_P+m_Q)^2}/{s}\Big)
\big(1-{(m_P-m_Q)^2}/{s}\Big)}\theta(s-(m_P+m_Q)^2)\ .
\en
The last term in eq.~\rf{UnitT01} accounts for isospin violation induced by
the $K^+-K^0$ mass difference via the function
\be
\Delta\sigma_K(s)=\frac{1}{2}\left(
\sigma_{K^+K^-}(s)- \sigma_{K^0\bar{K}^0}(s)
\right)\ .
\en
The unitarity relation for the $\bm{T}^{(12)}$ amplitudes, now, reads
\be
\im[ {\bm{T}^{(12)}} ]=  
\bm{T}^{(1)*}\,\Sigma^1\,{\bm{T}^{(12)}}
+{\bm{T}^{(12)*}}\sigma_{\pi\pi}\, T^{(2)}\ .
\en

\subsection{Coupled-channel KT equations}
The next step is to separate each isospin violating partial-wave amplitude,
which is an analytic function with two cuts into two functions, one
having a right-hand cut and one having a left-hand cut. In matrix form,
\be
\bm{T}^{(01)}=\frac{\sqrt6\,\epsilon_L}{32\pi}\left(
\bm{M}_0+\hat{\bm{M}}_0\right),\quad
\bm{T}^{(12)}=-\frac{\epsilon_L}{16\pi}\begin{pmatrix}
M_2+\hat{M}_2\\
G_{12}+\hat{G}_{12}\\
\end{pmatrix}\ .
\en
It is now easy to derive a matrix generalisation of the KT equations~\rf{KTeq}
involving $M_0$ and $M_2$. For the matrix $\bm{M}_0$, one has
\be\lbl{KTmatrix0}
\bm{M}_0(w)=\bm{\Omega}_0(w)\left[
\bm{P}_0(w)+ w^2 \left(\hat{\bm{I}}_A(w) +\hat{\bm{I}}_B(w)
\right)\right]{}^t\bm{\Omega}_1(w)
\en  
where $\bm{\Omega}_I$ are Muskhelishvili-Omn\`es (MO) $2x2$ matrices,
which must be computed numerically from the $I=0,1$ $T$-matrices and
$\bm{P}_0$ is a matrix of polynomials involving 12 parameters. The
matrices of integrals, finally, read
\be\lbl{IhatA}
\hat{\bm{I}}_A(w)=-\frac{1}{\pi}\int_{4\mpid}^\infty
\frac{ds'}{(s')^2(s'-w)}\left(
\im[\bm{\Omega}_0^{-1}(s')]\, \hat{\bm{M}}_0(s') \,{}^t\bm{\Omega}_1^{-1}(s') +
\bm{\Omega}_0^{-1*}(s')\, \hat{\bm{M}}_0(s')\, \im[{}^t\bm{\Omega}_1^{-1}(s')]
\right)
\en
and
\be\lbl{IhatB}
\hat{\bm{I}}_B(w)=\frac{32}{\sqrt6\,\epsilon_L}\int_{4 m^2_{\Kp}}^\infty
\frac{ds'\,\Delta\sigma_K(s')}{(s')^2(s'-w)}
\bm{\Omega}_0^{-1*}(s')\, \bm{T}^{(0)*}(s')
\begin{pmatrix}0 & 0\\
0 & 1\\ \end{pmatrix}
\bm{T}^{(1)}(s')\,{}^t\bm{\Omega}_1^{-1}(s')
\en
they correspond to the two different types of contributions in the
unitarity relation for $\bm{T}^{(01)}$. Note that in the
$\hat{\bm{I}}_B$ integrals isospin violation is induced by the
physical $K^+-K^0$ mass difference. As in the one channel case, the
representation~\rf{KTmatrix0} ensure that the unitarity
relations~\rf{UnitT01} for $\bm{T}^{(01)}$ are satisfied. It also
ensures that each component in the matrix $\bm{M}_0$ satisfies a
twice-subtracted dispersion relation associated with its right-hand
cut,
\be\lbl{dispM0}
[\bm{M}_0]_{ij}(w)= \alpha_{ij}+ \beta_{ij} w 
+ \frac{w^2}{\pi} \int_{4\mpid}^\infty
\frac{ds'}{(s')^2(s'-w)}\disc[[\bm{M}_0]_{ij}(s')]\ .
\en
Verifying that eq.~\rf{dispM0} is satisfied is a good check of the numerical
implementation of the KT representation~\rf{KTmatrix0}. 

The analogous equations derived from the $\bm{T}^{(12)}$ amplitudes read
\be\lbl{KTmatrix2}
\begin{pmatrix} 
M_2(w) &\\ 
G_{12}(w)\\ \end{pmatrix}=
\Omega_2(w)\bm{\Omega}_1(w) \begin{pmatrix}
w\,\beta_2 + w^2(\gamma_2 +\hat{I}_2(w) ) \\
\alpha_2^K + w\,\beta_2^K +w^2( \gamma_2^K + \hat{I}_2^K(w) \\
\end{pmatrix}
\en
with
\be
\begin{pmatrix}
\hat{I}_2(w) ) \\
\hat{I}_2^K(w) \\
\end{pmatrix}=
-\frac{1}{\pi} \int_{4\mpid}^\infty
\frac{ds'}{(s')^2(s'-w)}\im[ \Omega_2^{-1} \bm{\Omega}_1^{-1}]
\begin{pmatrix}
\hat{M}_2(s')\\
\hat{G}_{12}(s')\\
\end{pmatrix}\ .
\en
Note that both the $f_0(980)$ and the $a_0(980)$ resonances are present in the
representation~\rf{KTmatrix0} of the $\bm{M}_0$ functions via the two
Omn\`es-Muskhelishvili matrices (see fig.~\fig{OMcomp}). 
 
\subsection{Matching equations}
\begin{figure}
\centering
\includegraphics[width=0.49\linewidth]{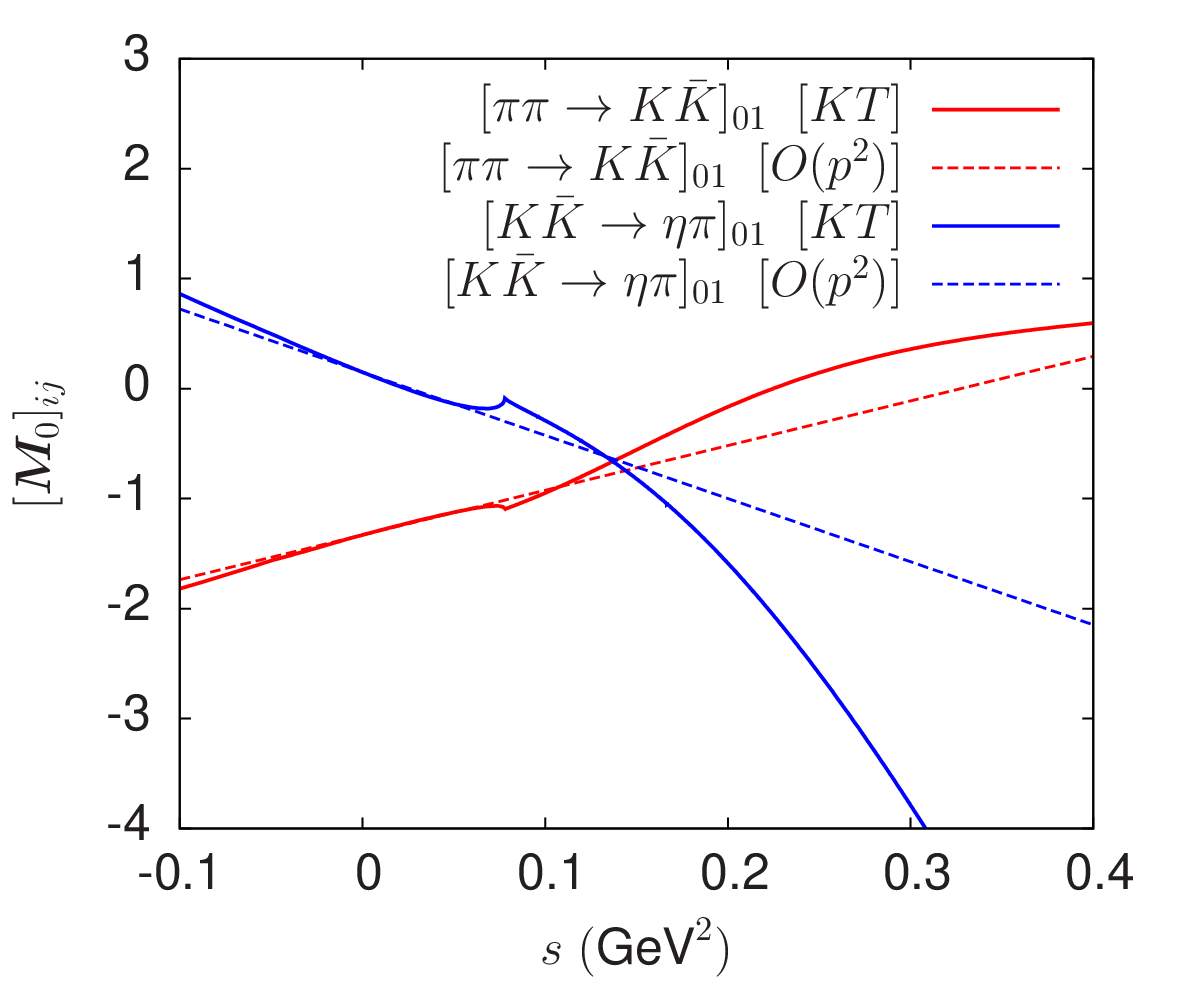}\includegraphics[width=0.49\linewidth]{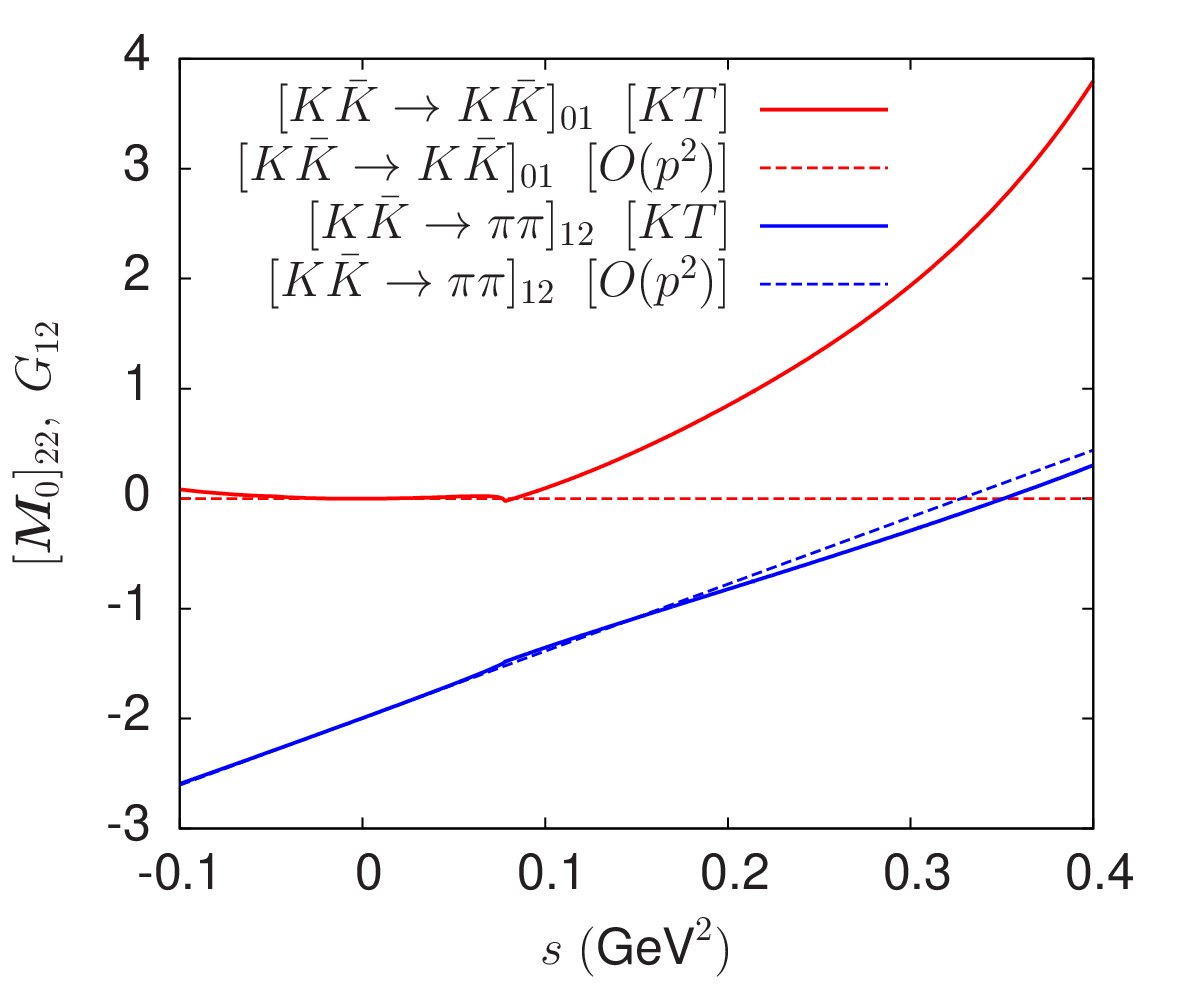}
\caption{$K\Kbar$ isospin violating amplitudes matched to the $O(p^2)$ chiral amplitudes}
\label{fig:KKamplits}
\end{figure}

For the $\eta\pi\to \pi\pi$ amplitudes we can use, as before, the four
matching equations associated with the NLO chiral amplitudes. The set of
matching equations~\rf{Matcheqs} is easily generalised to the coupled-channel
situation by using the corresponding KT representations for $M_0$, $M_1$,
$M_2$.
In contrast, for the amplitudes which involve the $K\Kbar$ channels, we do not
have a complete set of self-consistent KT equations which would enable us to
determine the left-cut functions in terms of the right-cut ones. Thus, we
cannot use the NLO amplitudes to perform the matching. The set of equations is
sufficient, however, for matching to leading order chiral amplitudes. Indeed,
at LO the chiral amplitudes have no left-hand cut. Accordingly, we can make
the approximation to set the left-cut functions $[\hat{\bm{M}}_0]_{ij}$ equal to
zero when $ij=12,21,22$ and $\hat{G}_{12}=0$. The polynomial parameters may
then be fixed such as to reproduce the $O(p^2)$ expressions
\be
\ba{ll}
[{\bm{M}}_0]_{21}=\dfrac{\sqrt6(3s-4\mkd)}{8(\mkd-\mpid)}\ ,\quad &
[{\bm{M}}_0]_{12}=-\dfrac{\sqrt3(3s-4\mpid)}{4(\mkd-\mpid)} \\[3pt]
[{\bm{M}}_0]_{22}=0\ , &
{G}_{12}=\dfrac{3\sqrt6(3s-4\mkd)}{16(\mkd-\mpid)}\ .\\
\ea
\en
The low energy behaviour of the these isospin
violating $K\Kbar$ amplitudes, resulting from this matching procedure, 
is illustrated on fig.~\fig{KKamplits}.
\section{Some results and comparisons with experiment}
\begin{figure}
\centering
\includegraphics[width=0.49\linewidth]{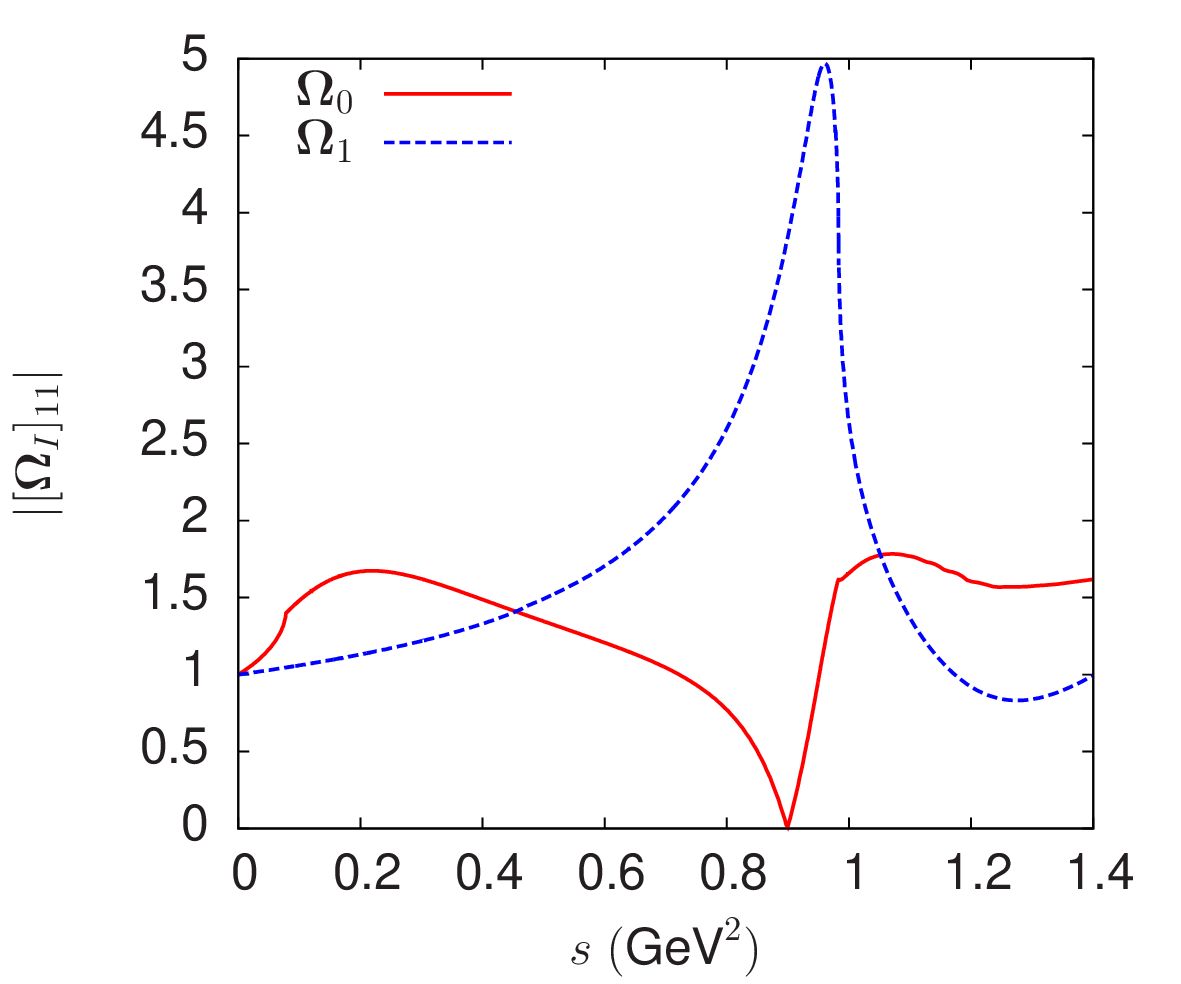}\includegraphics[width=0.49\linewidth]{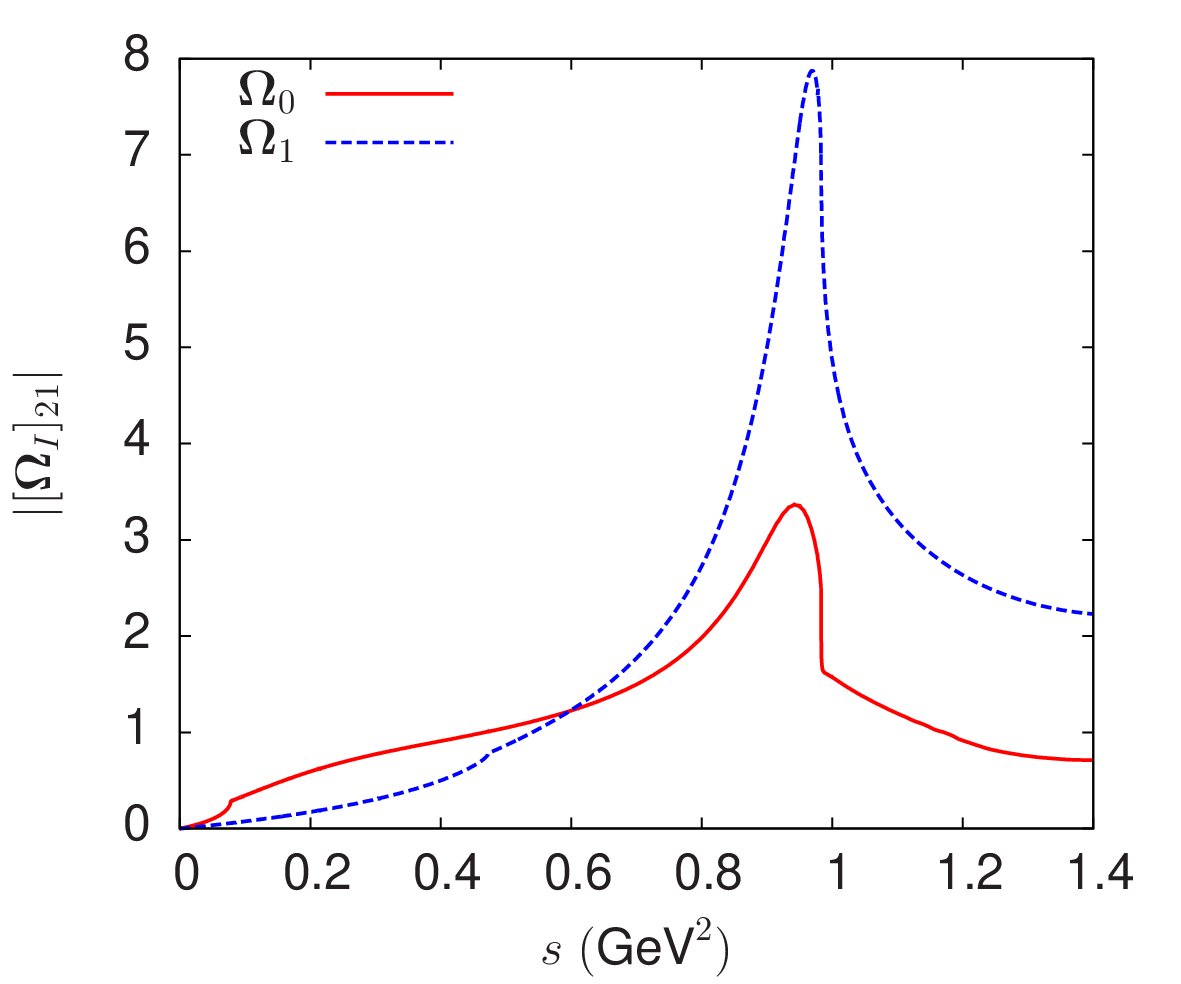}  
\caption{Some numerical results for the components of the MO matrices with
  $I=0$ and $I=1$.}
\label{fig:OMcomp}
\end{figure}
\subsection{Muskhelishvili-Omn\`es matrices $\bm{\Omega}_0$, $\bm{\Omega}_1$}
We still need to define the $I=0,\,1$ MO matrices which enter in the
formulation of the coupled-channel KT
equations~\rf{KTmatrix0}~\rf{KTmatrix2}. For $I=0$, we can rely on extensive
phase-shift analysis of $\pi\pi$ scattering performed long ago. New
high-precision measurements of the $S$-wave phase shift near threshold from
$K_{l4}$ decays have appeared~\cite{Batley:2010zza} and new Roy equations
solutions have been derived~\cite{Ananthanarayan:2000ht,GarciaMartin:2011cn}.
Measurements of the  inelastic $\pi\pi\to K\Kbar$ amplitude have also
been performed (see e.g.~\cite{Martin:1976mb}). This allows one to derive 
a two-channel model for the $T$-matrix and then, imposing appropriate
asymptotic conditions,  to compute numerically the corresponding MO matrix 
$\bm{\Omega}_0$ by standard methods~\cite{Babelon:1976kv,Donoghue:1990xh}. For
$I=1$, in contrast, there have been no measurements of $\pi\eta$ scattering
phase-shifts, but the properties of the $a_0(980)$ have been established via
its final-state interaction effects. We will use here a two-channel $T$-matrix
model~\cite{Albaladejo:2015aca} constrained by the properties of the $a_0(980)$
and $a_0(1450)$ resonances and by matching with NLO ChPT at low energy. We
also used implemented NLO chiral results on $\eta\pi$ and $K\Kbar$ $I=1$ form
factors, which are linearly related to the MO matrix $\bm{\Omega}_1$, and
provide additional constraints on the behaviour of the phase shifts above 1
GeV. As an illustration, we show in fig.~\fig{OMcomp} the modulus of the
components $[\bm{\Omega}_I]_{11}$, $[\bm{\Omega}_I]_{21}$ for $I=0,\,1$.

\subsection{$\eta\to 3\pi$ amplitudes from KT solutions}
\begin{figure}
\centering
\includegraphics[width=0.49\linewidth]{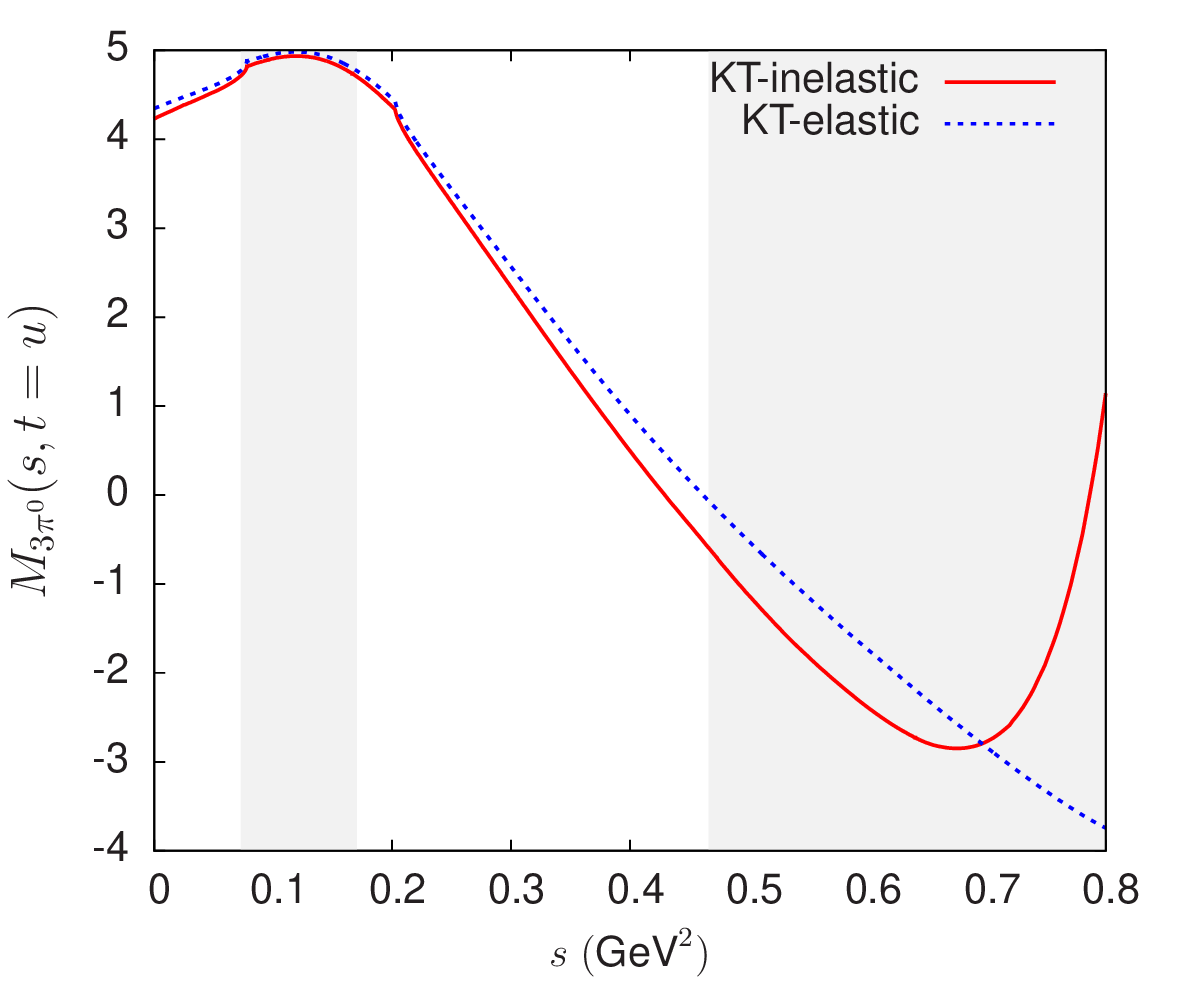}\includegraphics[width=0.49\linewidth]{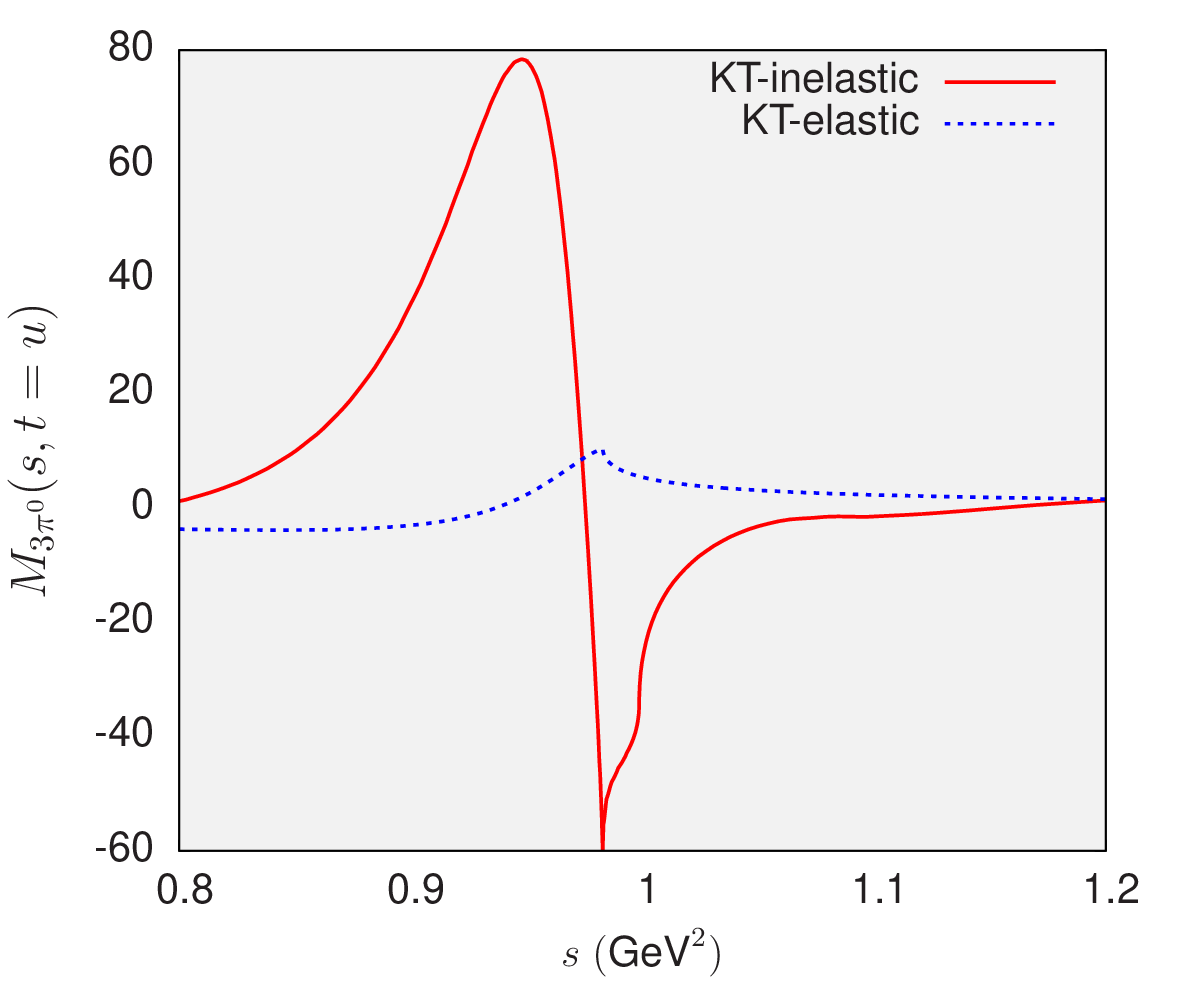}  
\caption{Real part of the amplitude $M_{3\pi^0}$ along the line $t=u$
  obtained from solving KT equations (dashed  line: elastic
  approximation, solid line: inelastic extension). The shaded areas
  indicate the physical regions of the decay $\eta\to 3\pi^0$ and the
  scattering $\eta\pi^0 \to \pi^0\pi^0$. }
\label{fig:M3pi0}
\end{figure}

Results for the $\eta\to 3\pi^0$ amplitude obtained from solving the
KT equations in the elastic approximation and their inelastic
extension as discussed above, are shown in fig.~\fig{M3pi0}. The
figure shows  that the  $a_0$ and $f_0$ resonances induce a very
large energy variation in the 1 GeV region, which is essentially
absent in the elastic approximation, and how this effect propagates
down to lower energies.

Numerical (preliminary) results for the Dalitz plot parameters of
$\eta\to \pi^+\pi^-\pi^0$ and $\eta\to \pi^0\pi^0\pi^0$ are shown in
table~\Table{Dalitz} and compared with experimental results. The
results associated with KT solutions are predictions based on the
matching equations~\rf{Matcheqs} and involve no fitted parameter. The
improvement, as compared to using the chiral NLO amplitude directly in the
physical region is spectacular\footnote{We agree on this point with the
results of ref.~\cite{Kambor:1995yc} but not with~\cite{Lanz:2011} who made
some approximations when implementing the matching conditions.}.  The
effects of the $a_0,f_0$ resonances can be as large as 10\% and seem to
improve the agreement with experiment, in particular for the parameter
$\alpha$.

\section{Conclusions}
We have developed a formalism which, within a simple approximation scheme,
allows us to take into account the influence of the scalar resonances
$a_0(980)$, $f_0(980)$ as well as the $K^+-K^0$ mass difference in the
$\eta\to 3\pi$ amplitude within the dispersive framework of Khuri and
Treiman. Matching with the chiral $O(p^4)$ amplitude in the sub-threshold
region, a prediction for the Dalitz plot parameters is achieved which agrees
reasonably well with experiment and this agreement is improved when the scalar
resonances are introduced. These results suggest that residual $O(p^6)$
effects (i.e. which cannot be ascribed to final-state interactions) could be
relatively small. An estimate of these effects should, however, be useful for
deriving a precise  value of the double quark mass ratio Q.
 
\begin{table}[h]
\centering
\fbox{$\eta\to \pi^+\pi^-\pi^0$}\\[2pt]
\bt{c||c|c|c||c||c}\hline\hline
Param. & $O(p^4)$ & KT-elastic & KT-coupled & WASA   & KLOE \\
\hline
a & -1.320 & -1.154 & -1.146 & -1.144(18) & -1.090(14) \\
\hline
b &  0.422 &  0.202 &  0.181 &  0.219(19) &  0.124(11) \\
\hline
f &  0.015 &  0.107 &  0.116 &  0.115(37) &  0.140(20) \\
\hline
d &  0.083 &  0.088 &  0.090 &  0.086(18) &  0.057(17) \\
\hline\hline
\et
\vspace{4pt}\\
\fbox{$\eta\to \pi^0\pi^0\pi^0$}\\[2pt]
\bt{c||c|c|c||c}\hline\hline
Param. & $O(p^4)$ & KT-elastic & KT-coupled & PDG  \\
\hline
$\alpha$ & +0.014 & -0.027 & -0.031 & -0.0315(15)  \\
\hline\hline
\et
\caption{Dalitz plot parameters from solutions of elastic and
  inelastic KT equations (preliminary results) with NLO chiral matching,
  compared to experimental determinations}
\lbltab{Dalitz}
\end{table}

\begin{acknowledgments}
This research was supported by Spanish Ministerio de Econom\'ia y
Competitividad and European FEDER funds under contracts FIS2014-51948-C2-1-P,
FPA2013-40483-P and FIS2014-57026-REDT, and by the European Community-Research
Infrastructure Integrating Activity "Study of Strongly Integrating Matter"
(acronym HadronPhysics3, Grant Agreement Nr 283286) under the Seventh
Framework Programme of the EU.  M.~A.~acknowledges financial support from the
''Juan de la Cierva'' program (reference 27-13-463B-731) from the Spanish
Government through the Ministerio de Econom\'{\i}a y Competitividad.
\end{acknowledgments}

\bibliographystyle{epj}
\bibliography{essai,uchpt,ffactor,revtex-custom}

\begin{thebibliography}{22}

\bibitem{Ambrosino:2008ht}
F.~Ambrosino et~al. (KLOE), JHEP \textbf{0805}, 006 (2008), \texttt{0801.2642},
%
P.~Adlarson et~al. (WASA@COSY), Phys.Rev. \textbf{C90}(4), 045207 (2014),
  \texttt{1406.2505}

\bibitem{Adolph:2008vn}
C.~Adolph et~al. (WASA@COSY), Phys.Lett. \textbf{B677}, 24 (2009),
  \texttt{0811.2763}, 
%
M.~Unverzagt et~al. (Crystal Ball@MAMI , TAPS , A2), Eur.Phys.J.
  \textbf{A39}, 169 (2009), \texttt{0812.3324}, 
%
S.~Prakhov et~al. (Crystal Ball@MAMI, A2), Phys.Rev. \textbf{C79}, 035204
  (2009), \texttt{0812.1999}, 
%
F.~Ambrosino et~al. (KLOE), Phys.Lett. \textbf{B694}, 16 (2010),
  \texttt{1004.1319}

\bibitem{Gasser:1984pr}
J.~Gasser, H.~Leutwyler, Nucl.Phys. \textbf{B250}, 539 (1985)

\bibitem{Bijnens:2007pr}
J.~Bijnens, K.~Ghorbani, JHEP \textbf{0711}, 030 (2007), \texttt{0709.0230}

\bibitem{Neveu:1970tn}
A.~Neveu, J.~Scherk, Annals Phys. \textbf{57}, 39 (1970)

\bibitem{Roiesnel:1980gd}
C.~Roiesnel, T.N. Truong, Nucl. Phys. \textbf{B187}, 293 (1981)

\bibitem{Kambor:1995yc}
J.~Kambor, C.~Wiesendanger, D.~Wyler, Nucl.Phys. \textbf{B465}, 215 (1996),
  \texttt{hep-ph/9509374}

\bibitem{Anisovich:1996tx}
A.~Anisovich, H.~Leutwyler, Phys.Lett. \textbf{B375}, 335 (1996),
  \texttt{hep-ph/9601237}

\bibitem{Khuri:1960zz}
N.~Khuri, S.~Treiman, Phys.Rev. \textbf{119}, 1115 (1960)

\bibitem{Achasov:1979xc}
N.~Achasov, S.~Devyanin, G.~Shestakov, Phys.Lett. \textbf{B88}, 367 (1979)

\bibitem{Batley:2010zza}
J.~Batley et~al. (NA48-2), Eur.Phys.J. \textbf{C70}, 635 (2010)

\bibitem{Ananthanarayan:2000ht}
B.~Ananthanarayan, G.~Colangelo, J.~Gasser, H.~Leutwyler, Phys. Rept.
  \textbf{353}, 207 (2001), \texttt{hep-ph/0005297}

\bibitem{GarciaMartin:2011cn}
R.~Garcia-Martin, R.~Kaminski, J.R. Pelaez, J.~Ruiz~de Elvira, F.J. Yndurain,
  Phys. Rev. \textbf{D83}, 074004 (2011), \texttt{1102.2183}

\bibitem{Martin:1976mb}
B.R. Martin, D.~Morgan, G.~Shaw, \emph{{Pion Pion Interactions in Particle
  Physics}} (Academic press, London, 1976)

\bibitem{Babelon:1976kv}
O.~Babelon, J.L. Basdevant, D.~Caillerie, G.~Mennessier, Nucl.Phys.
  \textbf{B113}, 445 (1976)

\bibitem{Donoghue:1990xh}
J.F. Donoghue, J.~Gasser, H.~Leutwyler, Nucl.Phys. \textbf{B343}, 341 (1990)

\bibitem{Albaladejo:2015aca}
M.~Albaladejo, B.~Moussallam (2015), \texttt{1507.04526}

\bibitem{Lanz:2011}
S.~Lanz, Ph.D. thesis, University of Bern (2011)

\end{thebibliography}
\end{document}